\documentclass[prd, twocolumn, amsfonts, amssymb, amsmath, showkeys, nofootinbib, superscriptaddress]{revtex4-2}

\usepackage{amsmath,amssymb,amsfonts}
\usepackage{graphicx}
\usepackage{dcolumn}       % align table columns on decimal point
\usepackage{bm}            % bold math
\usepackage{hyperref}
\usepackage{xcolor}
\usepackage{physics}       % \dd, \bra, \ket, \abs, etc.
\usepackage{siunitx}       % \SI{3.0e8}{\meter\per\second}
\usepackage{enumitem}
\usepackage{orcidlink}
\usepackage{newtxtext}
\usepackage{newtxmath}
\hypersetup{
  colorlinks = true,
  linkcolor  = blue,
  citecolor  = blue,
  urlcolor   = blue
}

\begin{document}

% ------------------------------------------------------------
% TITLE / AUTHORS / AFFILIATIONS
% ------------------------------------------------------------
\title{One-Percent Measurement of $H_0$ from Dark Sirens with Targeted Spectroscopic Follow-Up}

\author{Yixuan Dang\orcidlink{0000-0002-6689-8680}}
\email{ykd5167@psu.edu}
\affiliation{Institute for Gravitation and the Cosmos, The Pennsylvania State University, University Park, PA 16802, USA}
\affiliation{Department of Physics, The Pennsylvania State University, University Park, PA 16802, USA}

\author{Ariel J. Amsellem\orcidlink{0000-0003-3433-2698}}
\affiliation{McWilliams Center for Cosmology and Astrophysics, Department of Physics, Carnegie Mellon University, Pittsburgh, PA 15213, USA}

\author{Ignacio Magaña Hernandez\orcidlink{0000-0003-2362-0459}}
\affiliation{McWilliams Center for Cosmology and Astrophysics, Department of Physics, Carnegie Mellon University, Pittsburgh, PA 15213, USA}

\author{Antonella Palmese\orcidlink{0000-0002-6011-0530}}
\affiliation{McWilliams Center for Cosmology and Astrophysics, Department of Physics, Carnegie Mellon University, Pittsburgh, PA 15213, USA}

\author{B. S. Sathyaprakash\orcidlink{0000-0003-3845-7586}}
\affiliation{Institute for Gravitation and the Cosmos, The Pennsylvania State University, University Park, PA 16802, USA}
\affiliation{Department of Physics, The Pennsylvania State University, University Park, PA 16802, USA}
\affiliation{Department of Astronomy and Astrophysics, The Pennsylvania State University, University Park, PA 16802, USA}

%% Add more authors following the same pattern.

%% ── Abstract (≤250 words for ApJL) ───────────────────────────────────────────
\begin{abstract}
Gravitational-wave events from compact binary coalescences can be used as standard sirens: they encode the luminosity distance to their sources, which yields a measurement of the Hubble constant ($H_0$) if the source's redshift is known. In the absence of an electromagnetic counterpart, $H_0$ can still be inferred statistically from the cataloged galaxies within the event's sky localization volume via the dark siren, or galaxy catalog method. In this work we examine how the depth of a volume-limited galaxy catalog affects this inference and identify the most efficient spectroscopic strategy for an unbiased measurement of $H_0$ within a median precision of $0.94\%$. The strategy is to carry out spectroscopic surveys to a depth of $r \sim 19.5$ for the ten best-localized events observed in one year by the LIGO Hanford, LIGO Livingston, and LIGO-India network at A$^\#$ sensitivity, where A$+$ and A$^\#$ denote successive upgrades to Advanced LIGO designed to achieve progressively greater gravitational-wave sensitivity. Such a campaign requires a minimum of one night of electromagnetic follow-up observations, or five nights under the most conservative assumptions. For the Hanford, Livingston, and Virgo network at A$+$ sensitivity, the same ten-event strategy yields a median precision of $2.96\%$.
\end{abstract}

\maketitle
\textit{Introduction}---The Hubble constant $H_0$, which sets the present-day expansion rate of  the Universe, is among the most fundamental parameters in cosmology, yet its  value remains uncertain. Analyses of the cosmic microwave background anisotropies within the standard flat $\Lambda$CDM model yield $H_0 = 67.4 \pm 0.5\,\mathrm{km\,s^{-1}\,Mpc^{-1}}$ \citep{Planck:2018vyg}, while local distance-ladder measurements based on Cepheid-calibrated Type~Ia supernovae give $H_0 =73.2 \pm 0.9\,\mathrm{km\,s^{-1}\,Mpc^{-1}}$ \citep{Riess:2021jrx,Riess:2025chq}. The discrepancy between these two determinations now stands at $\sim$5$\sigma$ \citep{Verde:2019ivm,CosmoVerseNetwork:2025alb}, a tension that has resisted increasingly precise measurements and motivated intensive scrutiny of both known systematics and possible extensions to standard cosmology \citep{Poulin:2018cxd,DiValentino:2019qzk,DiValentino:2021izs,CosmoVerseNetwork:2025alb}. %Independent probes, including strong gravitational lensing time delays \citep{Wong2020}, the tip of the red giant branch \citep{Freedman2025}, and baryon acoustic oscillations \citep{DESICollaboration2024}, have so far failed to converge on a single resolution. A measurement of $H_0$ that is anchored to fundamentally different physics---one that bypasses the rungs of the distance ladder and the assumptions of early-Universe physics simultaneously---is therefore of great scientific value.

Gravitational-wave (GW) standard sirens offer an independent route to understanding this tension \citep{Schutz:1986gp,Holz:2005df,2026enap....5..557P}. Compact binary coalescences enable self-calibrated luminosity distance measurements from the GW waveform alone, without requiring external calibrators. When an electromagnetic (EM) counterpart is identified and the host galaxy redshift is measured directly, the combination yields an $H_0$ constraint from a single event, the so-called \emph{bright siren} method, first demonstrated with GW170817 and its optical counterpart \citep{abbott_gravitational-wave_2017,Palmese:2023beh}. However, binary black hole (BBH) mergers, which dominate the LIGO--Virgo--KAGRA (LVK) detection rate, produce no detectable EM emission in most scenarios. For these \emph{dark sirens}, $H_0$ and other cosmological parameters can be instead inferred by statistically associating the GW localization volume with galaxies in a redshift catalog \citep{Schutz:1986gp,DelPozzo:2011vcw,Mastrogiovanni:2023emh,Palmese:2021mjm, Palmese:2023beh,gair_hitchhikers_2023, Gray:2021thesis, Gray:2019ksv}. The current GWTC-5 catalog yields $H_0 = 71^{+9.0}_{-7.1}\,\mathrm{km\,s^{-1}\,Mpc^{-1}}$ \citep{LIGOScientific:2026uyd} with both bright and dark sirens, which has not yet converged to a resolution useful to the Hubble tension. 

There are other methods by which one can infer $H_0$ without relying on an EM counterpart. The method of \emph{spectral sirens} \citep{Taylor:2012db, Mastrogiovanni:2021wsd, Ezquiaga:2022zkx,Mali:2024wpq,Mukherjee:2021rtw} infers $H_0$ using features in the mass spectrum of compact binaries to break the mass-redshift degeneracy, which has provided a constraint of $H_0 = 70.1^{+15.1}_{-13.3}\,\mathrm{km\,s^{-1}\,Mpc^{-1}}$ alone \citep{LIGOScientific:2026uyd}. Other methods include the \emph{stochastic sirens} \citep{Cousins:2025bas}, \emph{Love sirens} \citep{Messenger:2011gi, Chatterjee:2021xrm, Dhani:2022ulg}, and cross-correlation of GW sources with the large-scale distribution of galaxies \citep{Namikawa:2015prh, Mukherjee:2019wcg,Mukherjee:2020hyn,Bera:2020jhx, Mukherjee:2022afz,Afroz:2024joi,Ghosh:2023ksl,Ghosh:2025qwc,Cheng:2026atn,Diaz:2021pem,Cross-Parkin:2026wyz}.

The constraint on $H_0$ obtained from the dark siren method scales inversely with the number of plausible host galaxies in the GW localization volume, which in turn depends on the sky localization area, the distance uncertainty, and the depth and completeness of the galaxy redshift catalog. Current analyses rely primarily on GLADE+ \citep{Dalya:2018cnd,Dalya:2021ewn}, Dark Energy Survey (DES \cite{DES:2005dhi,DES:2016jjg}) \cite{McMahon:2026nhi} which are both used in production of GWTC-5.0 cosmological constraints\citep{LIGOScientific:2026uyd}. Another work \cite{Palmese:2021mjm} has made use of the Dark Energy Spectroscopic Instrument (DESI \cite{DESI:2016fyo}) Legacy Survey. These catalogs provide mostly photometric redshifts, and can be highly incomplete in either depth (GLADE+) or sky coverage (DES). Replacing photometric redshifts with spectroscopic measurements (see, e.g., Ref.\,\cite{DESI:2023fij}) provides a further improvement. Photometric uncertainties of $\sigma_z \sim 10^{-2}$ are one to two orders of magnitude larger than the typical spectroscopic precision of $10^{-3}$--$10^{-4}$, depending on the source and instrument. For well-localized events, spectroscopic redshifts have been shown to improve constraints on $H_0$ by up to 15\% \citep{cross-parkin_dark_2025}.

Recent work has highlighted a particularly compelling subclass of dark sirens: events for which the sky localization is tight enough that the GW volume contains only a handful of galaxies, providing highly informative $H_0$ posteriors. Refs.\, \cite{Borhanian:2020vyr} and \cite{Gupta:2022fwd} identified these as \emph{golden dark sirens}, and \cite{Dang:2025vqx} demonstrated that a combined sample of golden (localized to within $0.1\,\mathrm{deg}^2$) and silver (localized to within $1\,\mathrm{deg}^2$) dark sirens observed over a single year can yield a few-percent constraint on $H_0$, provided that a deep, complete spectroscopic catalog is available within the localization volume, motivating dedicated spectroscopic follow-up campaigns for well-localized GW events.

A key question for any such spectroscopic campaign is how deep the galaxy catalog must be. Many analyses adopt a volume-limited sample, retaining only galaxies brighter than some fixed absolute luminosity threshold \citep[e.g.,][]{LIGOScientific:2018gmd,DES:2019ccw,DES:2020nay}. This choice is motivated by two considerations. First, a magnitude-limited sample is biased toward intrinsically bright galaxies at large distances, introducing a redshift-dependent selection effect that must be carefully modeled; a volume-limited sample, by construction, contains a complete and unbiased representation of galaxies above the luminosity threshold at all redshifts within the survey volume, simplifying the selection function. Second, BBH merger rates are expected to trace stellar mass or star formation rate, so that the true host galaxy is assumed to be more luminous than the chosen threshold. However, applying a bright luminosity cut, equivalent to a shallow apparent magnitude limit, also reduces the total number of galaxies in the catalog, which can artificially narrow the $H_0$ posterior and produce \emph{overconfident} constraints if the selection function is not perfectly modeled. Conversely, using a very deep catalog introduces many faint galaxies that contribute negligible prior weight but increase the observational cost of the inference. The optimal depth for a volume-limited catalog has not been systematically studied for well-localized events. 

Recent work in Ref.\,\cite{VanWyngarden:2025ogy} provides key foundations for the present study. Using mock catalogs drawn from MICECAT, they found that for well-localized events, retaining only the top 1\% of the top brightest galaxies is sufficient to recover an unbiased posterior as a direct consequence of galaxy clustering. They further suggest that restricting to spectroscopic catalogs may be preferable for future analyses to avoid biases from photometric redshift errors. Our work builds on these results in two directions. First, we determine the minimum apparent magnitude limit ($r\leq19.5$) at which the catalog does not underestimate the $H_0$ uncertainty relative to deeper surveys reaching $r\sim24$, since for a measurement intended to address the Hubble tension, a reliable uncertainty is crucial. Second, we translate this depth into a concrete observational program, quantifying the telescope resources required to achieve at the sub-percent measurement of $H_0$.

In this paper we address both questions for well-localized dark sirens, the optimal depth of the galaxy catalog and the observational cost of achieving complete spectroscopic coverage, in the context of the upgraded A$^\#$ \citep{Asharp} and A+ \cite{Aplus} networks. We make three concrete claims:
\begin{itemize}[nosep, leftmargin=18pt, itemindent=0pt]
    \item[(i)] The use of a shallow, volume-limited catalog with an apparent magnitude limit of $r \geq 19.5$ does not underestimate the uncertainty in $H_0$ for well-localized events compared to a much deeper catalog reaching $r \sim 24$, for samples of $ N\geq 10$ best well-localized events.
    \item[(ii)] Combining the 10 best-localized events from a single year of observations yields an unbiased (within the statistical uncertainty) measurement of $H_0$ to $0.94^{+0.49}_{-0.37}\%$ with the A$^\#$ network and to $2.96^{+1.74}_{-1.06}\%$ with the A$+$ network.
    \item [(iii)] Such an observational campaign with the A$^\#$ network  requires minimally 1 night, and more conservatively 5 nights, of follow-up telescope observations, making it feasible within several semesters' allocation at any of the facilities considered.
\end{itemize}

%% ═════════════════════════════════════════════════════════════════════════════
\textit{Statistical Framework for Dark Sirens}---Bayesian inference sets the statistical framework of our dark siren analysis. Given a collection of $N_{\mathrm{GW}}$ gravitational-wave event strain data denoted by $\{x\}$, and a galaxy catalog denoted as $\mathrm{CAT}$, the $H_0$ posterior can be written as:
\begin{equation}
p\left(H_0 \mid \{x\}, \mathrm{CAT}\right) \propto \mathcal{L}\left(\{x\} \mid H_0, \mathrm{CAT}\right) \times \pi(H_0),
\label{eq:H0posterior}
\end{equation}
where $\mathcal{L}$ denotes the joint likelihood of the GW data for the set of dark sirens, and $\pi(H_0)$ is the prior on the Hubble constant, taken to be uniform in the interval $[60, 80]~\mathrm{km\,s^{-1}\,Mpc^{-1}}$.

For the data of each GW event $x^{(i)}$, the joint likelihood over all $N_\mathrm{GW}$ events expands as \citep{Gray:2023wgj}:
\begin{eqnarray}
        \mathcal{L}(\{x\}\mid H_0,\mathrm{CAT}) 
        & \propto & \prod^{N_\mathrm{GW}}_i \int dz\, d\Omega\; \frac{\mathcal{L}(x^{(i)}\mid\Omega,\, d_{L}(z,H_0))}{\beta(H_0 \mid \mathrm{CAT}^{(i)}(H_0))} \nonumber \\ 
        & \times  & p(z,\Omega\mid H_0,\, \mathrm{CAT}^{(i)}(H_0))
\label{eq:bayes_extended_simplified}
\end{eqnarray} 
where $z$ is the redshift, $\Omega$ is the sky direction, and $d_L(z,H_0)$ is the luminosity distance computed assuming a flat $\Lambda$CDM cosmology with fixed matter density parameter $\Omega_m = 0.3$. We denote the volume-limited catalog of event $(i)$ by $\mathrm{CAT}^{(i)}(H_0)$. The single-event likelihood $\mathcal{L}(x^{(i)}\mid\Omega, d_L)$ encodes the probability of observing the GW detector response given a source at sky position $\Omega$ and luminosity distance $d_L$. The term  $p(z,\Omega\mid H_0,\, \mathrm{CAT}^{(i)}(H_0))$ characterizes how likely a galaxy is a host given a catalog as delta functions at the coordinates of galaxies included in the catalog, which also encodes the weighting scheme based on galaxy properties if any. As we only consider catalogs with well-measured spectroscopic redshifts, the redshift uncertainties are ignored. 

The denominator $\beta(H_0)$ is a normalization factor correcting for GW detector selection effects, defined as:
\begin{eqnarray}
        \beta(H_0 \mid \mathrm{CAT}(H_0)^{(i)}) 
        & = &  \int dz\, d\Omega\; P_{\mathrm{det}}(d_L(z,H_0),\,\Omega) \nonumber \\
        & \times & p(z,\Omega\mid H_0,\, \mathrm{CAT}^{(i)}(H_0))
\label{eq:beta}
\end{eqnarray}
where $P_{\mathrm{det}}(d_L(z,H_0),\,\Omega)$ is defined as
\begin{eqnarray}
    P_{\mathrm{det}}(d_L(z,H_0),\,\Omega) = \nonumber \\ \mathcal{L}(\mathrm{SNR}>11,  
     \Delta \Omega_{90} & <\Delta \Omega_{\rm threshold}|\Omega,d_L(z,H_0))
\label{eqn:Pdet}
\end{eqnarray}
where SNR is the signal-to-noise ratio of the detection, $\Delta \Omega_{90}$ is the 90\% credible interval sky localization, and $\Delta \Omega_{\rm threshold} = 0.1$ deg$^2$ for golden dark sirens, and $1$ deg$^2$ for silver dark sirens. For the best localized $N$ events, we use the maximum $\Delta \Omega_{90}$ of the best localized $10\times N$ events in 10 years of observation for their selection function.

\textit{Uchuu and GW Injections}---We construct a simulated catalog from Uchuu-$\nu^2$GC \citep{Oogi2023}, a dark matter N-body simulation \citep{Ishiyama2021} with galaxy properties assigned according to the semi-analytic model outlined in \cite{Makiya2016} and \cite{Shirakata2019}. Among the many galaxy properties, stellar mass and metallicity are particularly noteworthy for our analyses. They are constructed from the full star-formation and chemical enrichment histories composed from the simulation's many time steps. The star-formation history is also combined with the stellar population synthesis model of \cite{Bruzual+Charlot_2003} to compute a spectral energy distribution and corresponding absolute magnitudes. This process results in an approximately linear relationship between absolute magnitude and the decadic logarithm of the stellar masses. To convert this simulated catalog of galaxies and their properties to an observational lightcone, we take the original $2000~\mathrm{Mpc \, h^{-1}}$ simulated cube and periodically replicate that box many times to create a larger region of space. We then place an observer at the center of this box, and calculate the redshift of each galaxy as it would be observed by the observer. We follow this procedure for four shells covering the ranges $z=\{[0.0, 0.1], [0.1, 0.2], [0.2, 0.3], [0.3, 0.5]\}$ produced from four snapshots of the N-body simulation at redshifts $z=\{0.05, 0.14, 0.25, 0.43\}$, stitching together the four shells to create a galaxy catalog that extends out to a redshift of $0.5$ and only include galaxies with i-band magnitude of 24.1. A more detailed description of this procedure can be found in Section 3.3 of \cite{DongPaez2024}, while the lightcone is described in more detail in \cite{amsellem_darksiren}.\par 

We draw $5 \times 49{,}152 = 245{,}760$ galaxies from this derived Uchuu catalog, taking five galaxies from each of the $N_{\rm pix} = 49{,}152$ HEALPix pixels ($N_{\rm side} = 64$), resulting in an uniform distribution on the whole sky. The draw is restricted to galaxies with stellar mass $M_\star \in [10^8, 10^{12}]\,M_\odot$ and is weighted linearly by stellar mass according to
\begin{equation}
    w = \frac{M_\star - M_{\star,\min}}{M_{\star,\max} - M_{\star,\min}},
\end{equation}
so that higher-mass galaxies within the allowed range are preferentially selected. We also repeat the analysis with host galaxies drawn without this stellar mass weighting (see below), and in both cases all catalog galaxies are weighted equally in the $H_0$ inference.  The stellar mass weight applied at this step primarily prevents dwarf galaxies from being hosts of BBH mergers in our analysis. We note here that this stellar mass weighting is only applied at injection, but not in the $H_0$ inference.

Each potential host is then assigned BBH properties. We adopt the \textsc{Broken Power Law + 2 Peaks} model for mass and Gaussian Component Spin model for spin as in Ref.\ \cite{LIGOScientific:2025pvj}. For the detector-frame redshift distribution of BBH mergers we use $R_d(z) = R_c(z)\,(1+z)^{-1}\,dV_c/dz$, where $R_c$ is the comoving merger rate and $V_c$ the comoving volume. We take the BBH formation rate to follow the Madau-Dickinson SFR history \citep{madau_dickinson_2014}, convolved with a $1/t_d$ time-delay distribution between $t_\mathrm{min} = 50$~Myr and $t_\mathrm{max} = 13.63$~Gyr (0.1 Gyr after the Big Bang). The resulting comoving rate can be well described by the Madau-Dickinson form \citep{Ng:2020qpk}
\begin{equation}
    R_c(z) = \phi_0 \, \frac{(1+z)^{\gamma}}{1 + \left(\dfrac{1+z}{1+z_p}\right)^{\kappa}}
    \label{eq:redshift dist}
\end{equation}
with $(z_p,\, \gamma,\, \kappa) = (1.75,\, 2.0,\, 4.6)$, and $\phi_0$ fixed so that the local merger rate density is $16.16\,\mathrm{Gpc}^{-3}\,\mathrm{yr}^{-1}$ \citep{LIGOScientific:2025pvj}.

The Right Ascension (RA) and Declination (Dec) distribution is uniform across the whole sky. To respect the redshift distribution already encoded in Uchuu, we use this rate evolution model only to determine the total number of BBH mergers expected in 30 years, rather than to redraw individual event redshifts. Concretely, after the initial Fisher-matrix calculation of GW parameters by \texttt{GWBENCH} \citep{Borhanian:2020ypi} for the full population of 245,760 injections, we randomly subsample the number of BBH mergers expected in 30 years according to the rate evolution model above, leaving the underlying redshift distribution of the sampled events unchanged.\footnote{We compared the forecast number of golden and silver dark sirens obtained using the redshift distribution of
Equation~\ref{eq:redshift dist} 
against Uchuu's native redshift distribution and found good agreement; the two treatments are therefore equivalent for the purposes of this work.} We then select the 300 best-localized events from the Fisher analysis in 30 years of observations. For most events, we use the mode-by-mode relative binning method~\citep{Gupta:2024bqn} for waveform IMRPhenomXPHM \citep{Pratten:2020ceb}. For events where their mode-by-mode ratio computation fails because of silent higher order modes, we use the public relative binning method \citep{relbin_bilby,Zackay:2018qdy,relbin_cornish,Gupta:2024bqn} to accelerate inference in bilby \citep{bilby_paper}. Then we reweight the likelihood for each sample with the ratio between relative-binning likelihood and the exact likelihood. Finally, we select the best localized 90, 150, 240, 270 and 298 events from this 30-year dataset, and these events are bootstrapped to provide 1000 realizations of the best 3, 5, 8, 9 and 10 localized events per year. For $N=10$ we exclude two outlier events events whose sky area are much larger than the rest. One year was chosen to be similar to the effective duration of a typical observing run. For A$^{\#}$ sensitivity we adopt a network of LIGO Hanford, Livingston, and LIGO-India, denoted HLI$^{\#}$ in later sections; for A+ sensitivity we use LIGO Hanford, Livingston, and Virgo, denoted HLV+. For the comparison runs, HLV+ (with stellar-mass weighting) and HLI$^{\#}$ without stellar-mass weighting of the host galaxies at injection, we use a 10-year data set to limit the computational cost.

% In Table.\,\ref{tab:expected_number}, we present the number of well-localized events detected per year with the average total sky area of the 10 best localized events.
% \begin{table}[h]
% \centering
% \begin{tabular}{c|c|c|c}
% \hline
% \hline
% Network & N ($\Omega_{90} < 0.1$) & N ($\Omega_{90} 
% < 1$) \\
% \hline
% HLV+ & 0.2 & 5.5  \\
% HLI\# &2.3 &54.8  \\
% \hline
% \hline
% \end{tabular}
% \caption{Expected number of GW events per year averaged over a 10-year period with sky area localization thresholds corresponding to golden and silver dark sirens. Notice that the Fisher matrix estimation of sky area deviates from more accurate bilby results in some cases.}
% \label{tab:expected_number}
% \end{table}

%% ── Example figure ────────────────────────────────────────────────────────────
%% figure*  spans both columns — typical for the main result figure in ApJL.
%% Use figure (no *) for a single-column inset.
%\begin{figure*}[ht!]
%\plotone{figures/fig1.pdf}   % replace with your figure filename
%\caption{Caption for Figure 1. Be descriptive --- captions should be
%  self-contained.
%  \label{fig:main}}
%\end{figure*}

%% ═════════════════════════════════════════════════════════════════════════════

\textit{Optimal Galaxy Catalog Depth}---A shallower volume-limited sample gains more precision from the assumption that the galaxy is brighter than the absolute magnitude cut, which is an assumption one might challenge, while a deeper sample's uncertainty is derived under a much weaker version of this assumption. The optimal catalog depth is therefore the point beyond which additional depth no longer changes the $H_0$ uncertainty, while the results remain unbiased. We examine whether a shallow volume-limited sample underestimates this uncertainty relative to the deepest practical catalog at $r \sim 24$.

% \begin{table}[!b] 
% \centering 
% \begin{tabular}{c|ccc}
% \hline \hline & \multicolumn{3}{c}{Mean fractional 68\% CI width [\%]} \\ \cline{2-4} Siren ($N_{\rm events}$) & $m_{\rm app} \leq 19.5$ & $m_{\rm app} \leq 21.5$ & $m_{\rm app} \leq 23.5$ \\ \hline Silver ($N=307$) & 15.87& 16.14& 23.5\\ Golden ($N=57$)  & 7.081& 7.53& 7.968\\ \hline \hline 
% \end{tabular}
% \caption{Growth of the per-event mean fractional 68\% credible-interval (CI) width of the $H_0$ posterior, $\Delta H_0 / H_0^{\rm med}$ with $\pm 1\sigma_{\rm SEM}$ error bars ($\sigma/\sqrt{n}$, the standard error on the mean), as a function of the apparent-magnitude limit $m_{\rm app,\,lim}$ imposed on the host-galaxy catalog. Each magnitude threshold is mapped to a per-event absolute-magnitude cut $M_{\rm abs}\le m_{\rm app,\,lim}-\mu(D_{L,95})$ and applied to the credible-region galaxy sample of each of $n=98$ silver dark-siren injections. Blue and red circles represent silver and golden dark sirens respectively.} 
% \label{tab:fracwidth_vs_mlim} \end{table}

\begin{figure*}
    \centering
    \includegraphics[width=1.00\linewidth]{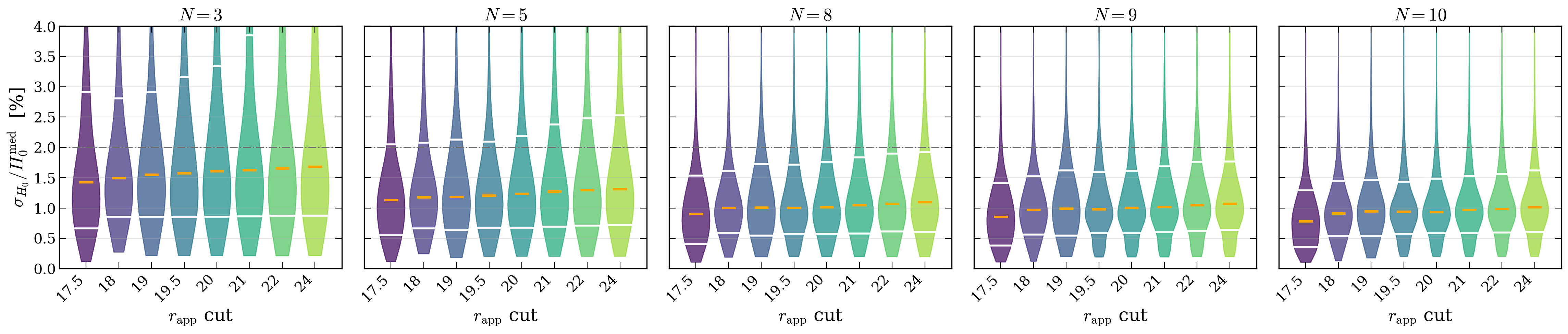}
    \caption{Fractional width $\sigma_{H_0}/H_0^{\rm med}$ of the joint $H_0$ posterior derived from the top 3, 5, 8, 9 and 10 best-localized dark sirens within one year of A$^\#$ operations, shown as a function of the catalog apparent-magnitude limit $r_{\rm app}$. For each cut and each $N$, $1000$ bootstrap realizations are drawn (each combining $N$ events sampled with replacement from $30\times N$ best localized events in the 30-year data), and the resulting distribution of $\sigma_{H_0}/H_0^{\rm med}$ is shown as a violin whose width is the probability density. The thick orange line marks the median over realizations and the thin white lines the $16$th and $84$th percentiles. The dash-dotted line marks the $2\%$ precision level.}
    \label{fig:width}
\end{figure*}

\begin{figure*}
    \centering
    \includegraphics[width=1.00\linewidth]{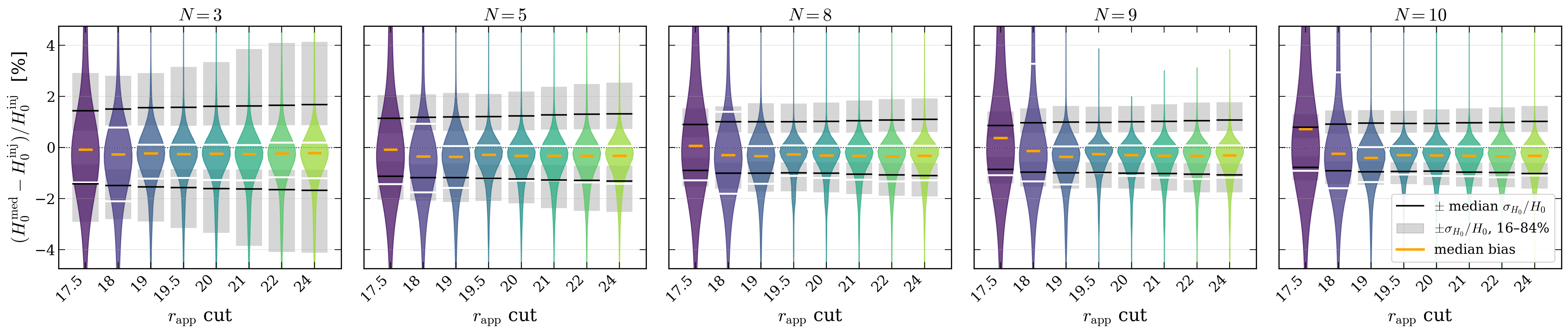}
    \caption{Fractional statistical fluctuation of the joint $H_0$ posterior median, $H_0^{\rm med}-H_0^{\rm inj}/H_0^{\rm inj}$ (with $H_0^{\rm inj}=67.74~{\rm km\,s^{-1}\,Mpc^{-1}}$) derived from the top 3, 5, 8, 9 and 10 best localized dark sirens within one year of A$^\#$ operations, shown as a function of the catalog apparent-magnitude limit $r_{\rm app}$. For each cut and $N$, the distribution of $H_0^{\rm med}-H_0^{\rm inj}$ over $1000$ bootstrap realizations (each combining $N$ events sampled with replacement from $30\times N$ best-localized events in the 30-year data) is shown as a violin whose width is the probability density; the thick orange line marks the median over realizations and the thin white lines the $16$th and $84$th percentiles. The dotted line marks zero. The black horizontal lines mark $\pm$ the median fractional uncertainty $\sigma_{H_0}/H_0$ across the bootstrap realizations and the grey bands span its 16th--84th percentile range, showing that the median bias lies well within the statistical uncertainty for every $N$ and magnitude cut.}

    \label{fig:stability}
\end{figure*}
\begin{figure}
    \centering
    \includegraphics[width=1.00\linewidth]{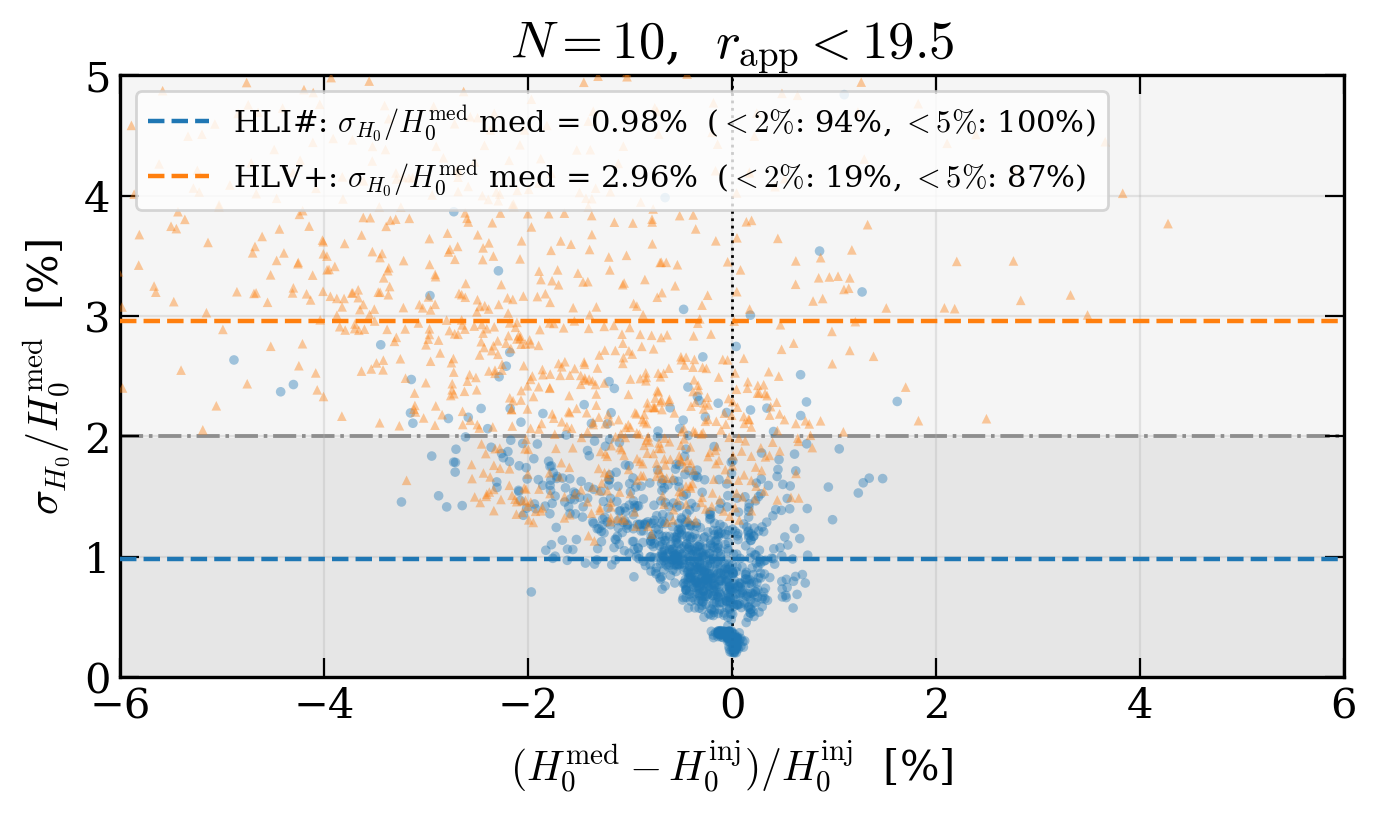}
    \caption{Summary of the $H_0$ measurement performance for the proposed strategy: the 10 best-localized events at a catalog depth of $r_\mathrm{app} < 19.5$, comparing the HLI\# (A\#) and HLV+ (A+) networks. Each point represents one of $1000$ bootstrap realizations, drawn with replacement from the pool of the best localized events. The horizontal axis shows the fractional offset $(H_0^\mathrm{med} - H_0^\mathrm{inj})/H_0^\mathrm{inj}$, while the vertical axis shows the precision $\sigma_{H_0}/H_0^\mathrm{med}$. Points falling within the shaded band achieve the $\sigma_{H_0}/H_0 < 2\%$ target. The dashed horizontal lines mark the median precision of each network. The HLI\# realizations cluster tightly near zero offset and well within the $2\%$ band. The HLV+ realizations show larger scatter in both axes, but $87\%$ of realizations still achieve $\sigma_{H_0}/H_0 < 5\%$, demonstrating that a meaningful few-percent constraint on $H_0$ is already attainable with the near-term A+ network before LIGO-India comes online.}
    \label{fig:2d width and fluc}
\end{figure}
Figure~\ref{fig:width} shows that deeper catalogs yield a slightly broader $H_0$ posterior for all values of $N$ considered, though the increase is modest: the median $\sigma_{H_0}/H_0$ differs by only $14\%-23\%$ between the shallowest ($r=17.5$) and deepest ($r\sim24$) catalogs for $N=3,5,8,9$ and $10$.
 For the 10 best-localized events, this difference becomes negligible: the median grows from $0.94\%$ for $r=19.5$ to $1.02\%$ for $r=24$, compared to the $2\%$ measurement goal.
We adopt $r\leq19.5$ because it is the shallowest cut at which the median $\sigma_{H_0}/H_0^\mathrm{med}=0.94\%$ tracks the spread of the posterior medians across realizations ($0.87\%$), whereas at $r\leq19$ the spread ($2.0\%$) is about twice the median $\sigma_{H_0}/H_0^\mathrm{med}$ ($0.95\%$), driven by a few events whose hosts fall below the cut and whose localization volumes contain only a handful of catalog galaxies. We noticed that several events are matched with no galaxies at all at $r \leq 18$, and an uninformative $H_0$ posterior is used for those events (8 of the 298 events at $r = 18$ and 31 at $r = 17.5$). When this occurs, the affected events contribute no cosmological information, reducing the effective number of informative dark sirens from $N$ to $N-1$ or even $N-2$.  Another category of interest comprises events whose host galaxy is removed by the $r \leq 19.5$ cut. There are 29 of the 298 events which yield posteriors built entirely from galaxies that are not the source. Galaxy clustering mitigates the effect, however, since even when the true host falls below the magnitude cut, neighboring bright galaxies at similar redshift provide support at approximately the correct redshift (see also~\cite{VanWyngarden:2025ogy}) As the figure shows, both the median and the precision of $H_0$ recovered at $r = 19.5$ are comparable to those at $r = 24$, where the host is present in every event.

% \begin{table*}[!b]
% \centering
% \begin{tabular}{c|cccccc}
% \hline
% \hline
% Telescope & DESI & Magellan & HET+VIRUS  & AAT&4most&PFS\\
% \hline
% Hours/GDS& ?& 0.1+0.5& 0.8+1.2& 1+1&2+0.5& 0.5+1.3\\
% Hours/SDS& ?& 0.6+0.5& 6.8+9.9& 3+1&2+0.5& 0.5+1.3\\
% \hline
% \hline
%  Hours/best-10& & & & & &\\
% \end{tabular}
% \caption{Estimation of telescope time required to follow-up well-localized events. Second and third rows stand for hours required per golden dark siren or silver dark siren. Last row stands for the hours required for the 10 best localized events. An estimate of $0.5+0.1$ stands for 0.5 hours for exposure, and 0.1 hours of overheads.}
% \label{tab:time}
% \end{table*}

\begin{figure*}
    \centering
    \includegraphics[width=0.95\linewidth]{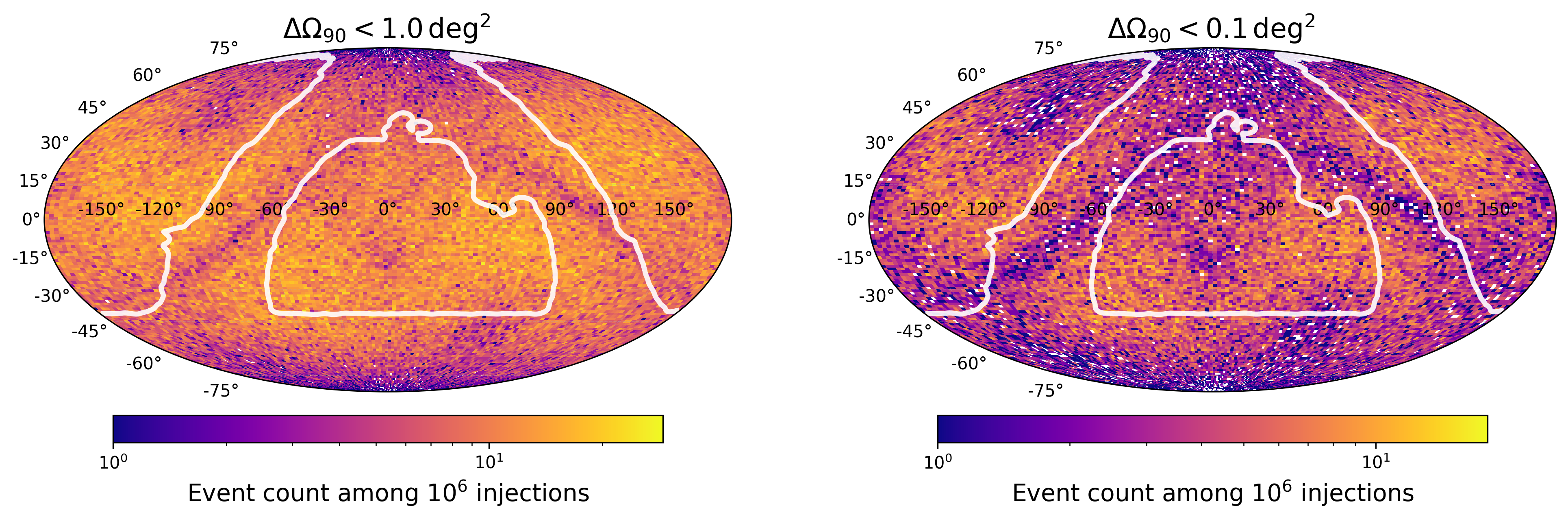}
    \caption{Sky location distribution map of silver (\emph{left}) and golden (\emph{right}) dark sirens. The white contours represent the DESI footprint. }
    \label{fig:map}
\end{figure*}

\textit{The Most Efficient Strategy}---Figures\,\ref{fig:width} and~\ref{fig:stability} together allow us to identify the most efficient strategy for an unbiased sub-$2\%$ measurement of $H_0$---a strategy that in fact delivers a median precision of $1\%$. Such a measurement demands two things at once: a forecast precision that falls below the $2\%$ target in the large majority of realizations, not merely the typical one, and a recovered $H_0$ that is statistically unbiased. The median $\sigma_{H_0}$ already falls below $2\%$ for the 3, 5, 8, 9 and 10 best-localized events. In particular, with the 10 best-localized events at $r=19.5$, $\sigma_{H_0}/H_0^\mathrm{med} = 0.94^{+0.49}_{-0.37}\%$, so even its $84$th percentile, $1.43\%$, lies safely below target and the median sits below $1\%$. The measurement is also unbiased at this depth: for the 10 best-localized events at $r = 19.5$, $(H_0^\mathrm{med} - H_0^\mathrm{inj})/H_0^\mathrm{inj} = -0.30^{+0.35}_{-0.75}~\%$ lies well within the statistical uncertainty. Even the $16$th and $84$th percentiles of the offset, $-1.06\%$ and $+0.05\%$, are comparable to the median precision of $0.94\%$, so the scatter of the recovered median across realizations is consistent with the forecast uncertainty. We adopt $N=10$ as a safety choice because, even at the deepest cut ($r=24$), where the posterior is broadest, the median $\sigma_{H_0}/H_0^\mathrm{med}$ first falls below $1\%$ at $N=9$ ($1.0\%$, compared with $1.02\%$ for $N=8$), and we include one additional event as a margin of safety. This choice does not have a significant impact on the telescope time as we see later.We suspect that the slight tilt of the median toward lower $H_0$ arises because the luminosity-distance medians are more often overestimated in our injections.  Finally, Figure~\ref{fig:2d width and fluc} shows that the forecast precision $\sigma_{H_0}/H_0^\mathrm{med}$ falls below the $2\%$ target in $95.4\%$ of the 1000 realizations. When aiming for the $2\sigma$ or $3\sigma$ level, the precision reached in $95.45\%$ ($2\sigma$) of realizations is $2.0\%$, and in $99.73\%$ ($3\sigma$) of realizations it is $\approx 5.1\%$, the latter being set by a handful of realizations in the tail of the distribution.
% , and the median $\sigma_{H_0}/H_0^\mathrm{inj}$ ($\approx1\%$) closely matches the run-to-run scatter of the recovered medians ($\approx1\%$). The measurement is therefore unbiased and statistically limited, and we conclude that following up the 10 best-localized events to $r \simeq 19$ is the most efficient strategy for a robust, unbiased $2\%$ determination of $H_0$.

When we switch to the LIGO Hanford, Livingston and Virgo network at A+ sensitivity \citep{Aplus}, the ten best-localized events yield $\sigma_{H_0}/H_0^\mathrm{med} =2.96^{+1.74}_{-1.06}\%$ at $r=19.5$. The larger sky-localization areas of this network result in only $19\%$ of the $1000$ realizations reaching the $2\%$ goal, with $87\%$ reaching $5\%$ or less. The median $\sigma_{H_0}/H_0^\mathrm{med}$ ($2.96\%$) still tracks the spread of recovered medians ($2.19\%$). As before, the choice of magnitude cut has little effect, with $\sigma_{H_0}/H_0^\mathrm{med}$ increased by only $0.08$ percentage points($\sim5\%$) between $r=19.5$ and $r=24$. 

Finally, one may ask whether the stellar-mass weighting applied when drawing injections drives this result; repeating the analysis of the 10 best-localized events detected by HLI\# at $r=19.5$ without the stellar-mass weighting at injection gives a slightly wider but still comparable $\sigma_{H_0}/H_0^\mathrm{med} = 1.24^{+0.73}_{-0.34}\%$. Without stellar-mass weighting the fractional posterior width at $r = 18 $ lies 15\% below its $r = 24$ value at $N = 10$, converging to within $\sim5\%$ only for $r = 19.5$.

We have identified a non-exhaustive list of telescopes and instruments worldwide suitable for conducting a spectroscopic follow-up survey of these well-localized events. In the Northern hemisphere, the Dark Energy Spectroscopic Instrument (DESI) \citep{DESI:2022xcl}, the Subaru Prime Focus Spectrograph (Subaru/PFS) \citep{Tamura:2016wsg,SubaruPFSInstrument}, and the Hobby--Eberly Telescope with the Visible Integral-field Replicable Unit Spectrograph (HET/VIRUS) \citep{Hill2021,het_descrip} are all capable of this task. In the Southern hemisphere, we have explored the Anglo-Australian Telescope (AAT) \citep{AAT2df}, the 4-metre Multi-Object Spectroscopic Telescope (4MOST) \citep{deJong2019}, and the Magellan Baade Telescope with the Inamori-Magellan Areal Camera and Spectrograph (Magellan/IMACS) \citep{imacs}.
While some facilities --- notably DESI, Magellan/IMACS and Subaru/PFS --- complete this follow-up substantially faster than others in one single night, instrument choice in practice also depends on sky location, scheduling availability, and access constraints. For example,  HET+VIRUS is markedly slower per event, but it is the only instrument considered here that requires no pre-existing photometric catalog, making it uniquely suited to localization regions lacking $r\geq19.5$ imaging coverage. On average, the time taken in total is $\sim 24.4$\,hr ($\lesssim 5$ nights, assuming 5 to 8 hours of usable dark time per night), sufficient to obtain a complete spectroscopic catalog for the 10 best-localized events in a typical observing year (see End Matter for details). In Figure \,\ref{fig:map}, we show the sky location distribution of golden and silver dark sirens, which indicates which regions are more likely to host a well-localized event. A large proportion of these regions are covered in the DESI footprint \citep{DES:2020nay}, which is expected to greatly reduce the new telescope resources required for this campaign thanks to available serendipitous observations.

%% ═════════════════════════════════════════════════════════════════════════════
\textit{Discussion and Conclusions} ---In this work we propose an efficient spectroscopic strategy for dark sirens: \emph{a dedicated spectroscopic follow-up campaign reaching an apparent magnitude no shallower than $r = 19.5$ for the 10 best-localized events detected in a year with HLI\#.} This campaign requires as little as one night of telescope time to obtain a spectroscopic catalog at the depth necessary. We also show that forfeiting stellar-mass weighting of host galaxies in the injections weakens the resulting $H_0$ constraint slightly. Several caveats remain. We have neglected the contribution of peculiar velocities to the redshift uncertainty, which introduces a typical additional error of $8 \times 10^{-4}$, increasing to at most $3 \times 10^{-3}.$ The origin of the small offset in the median of the recovered $H_0$ remains unclear; we suspect it is caused by a slight overestimation of the luminosity distance, but defer a detailed investigation to future work. We also find that roughly $20\%$ of these well-localized events fall within the zone of avoidance. Discarding them would extend the one-year observing period set to collect GWs by a comparable factor for the same results to hold.

%We have ignored the effects of photo-z uncertainties to the estimation of telescope time as described in \citet{Kaur:2025vre}.

%% ── Acknowledgments (not counted in word quota) ───────────────────────────────
\begin{acknowledgments}
We thank Robin Ciardullo and Joel Leja for helpful discussions on HET/VIRUS and Subaru/PFS. We also thank Zhuotao Li for the LIGO Scientific Collaboration internal review, Vaishak Prasad and Ish Gupta for improving the parameter estimation, and Utkarsh Mali for the comments. Y.D. is supported by National Science Foundation (NSF) grants AST-2307147 and NSF AST 2606673. B.S.S. is supported by NSF grants AST-2307147, AST 2606673, PHY-2308886, PHY-2309064, and PHY 2608011. A. P. is supported by NSF Grant No. 2308193. The authors would also like to acknowledge the LIGO Laboratory computing resources supported by NSF grants PHY-0757058 and PHY-0823459, and the Gwave cluster maintained by the Institute for Computational and Data Sciences at Penn State University, supported by NSF grants: OAC2346596, OAC2201445, OAC2103662, OAC2018299, and PHY-2110594. The Institute for Gravitation and the Cosmos is supported by the Eberly College of Science and the Office of the Senior Vice President for Research at the Pennsylvania State University. This work used resources on the Vera Cluster at the Pittsburgh Supercomputing Center. The authors used Claude (Anthropic) to assist with software development and analysis, and all outputs were reviewed by the authors.
\end{acknowledgments}

 \textit{Software:} \texttt{astropy} \citep{astropy2018}, \texttt{GWBENCH} \citep{Borhanian:2020ypi}, \texttt{bilby} \citep{bilby_paper}, \texttt{Matplotlib} \citep{2007matplotlib}, \texttt{NumPy} \citep{Harris_2020}, \texttt{SciPy} \citep{Virtanen_2020}
\appendix
\section{End Matter}

\begin{table*}
\centering
\caption{Spectroscopic follow-up requirements for the $10$ best-localized dark sirens in one year. Thirty (HLI\#) and ten (HLV+) years of events were simulated; each HLI\# rank is a block of 30 events and each HLV+ rank a block of 10, sorted by localization area, and the last HLI\# rank contains the remaining 28 events after the two outliers are excluded. The second and third columns give the mean $90\%$ localization area for HLI\# and for the HLV+ network for comparison. The fourth and fifth columns give the mean injected redshift of the same events for each network. The remaining columns list exposure and overhead times per facility, computed for the HLI\# localizations; the ``All'' row gives the summed follow-up time.}
\begin{tabular}{c|cc|cc||cccccc}
\hline\hline
 & \multicolumn{2}{c|}{Mean $\Delta \Omega_{90}$ [deg$^2$]} & \multicolumn{2}{c||}{Mean $z$} & \multicolumn{6}{c}{Exposure + Overhead (hours)} \\
\cline{2-3}\cline{4-5}\cline{6-11}
Rank & HLI\# & HLV+ & HLI\# & HLV+ & DESI & Magellan/IMACS & HET/VIRUS & AAT/2dF & VISTA/4MOST & Subaru/PFS \\
\hline
$1$  & $0.043$ & $0.287$ & $0.082$ & $0.112$ & $0.1+0.16$ & $0.1+0.1$ & $0.4+0.6$ & $4+1$ & $3.2+1.8$ & $0.5+0.3$\\
$2$  & $0.078$ & $0.472$ & $0.114$ & $0.141$ & $0.1+0.16$ & $0.1+0.1$ & $0.8+1.2$ & $4+1$ & $3.2+1.8$ &  $0.5+0.3$ \\
$3$  & $0.106$ & $0.611$ & $0.125$ & $0.187$ & $0.1+0.16$ & $0.1+0.1$ & $0.8+1.2$ & $4+1$ & $3.2+1.8$ & $0.5+0.3$ \\
$4$  & $0.134$ & $0.725$ & $0.155$ & $0.150$ & $0.1+0.16$ & $0.1+0.1$ & $1.2+1.8$ & $4+1$ & $3.2+1.8$ & $0.5+0.3$ \\
$5$  & $0.152$ & $0.806$ & $0.159$ & $0.162$ & $0.1+0.16$ & $0.1+0.1$ & $1.2+1.8$ & $4+1$ & $3.2+1.8$ & $0.5+0.3$ \\
$6$  & $0.176$ & $0.906$ & $0.152$ & $0.176$ & $0.1+0.16$ & $0.1+0.1$& $1.2+1.8$ & $4+1$ & $3.2+1.8$ & $0.5+0.3$ \\
$7$  & $0.198$ & $0.977$ & $0.147$ & $0.215$ & $0.1+0.16$ & $0.2+0.2$ & $1.6+2.4$ & $4+1$ & $3.2+1.8$ & $0.5+0.3$ \\
$8$  & $0.221$ & $1.070$ & $0.172$ & $0.169$ & $0.1+0.16$ & $0.2+0.2$ & $1.6+2.4$ & $4+1$ & $3.2+1.8$ & $0.5+0.3$ \\
$9$  & $0.252$ & $1.159$ & $0.207$ & $0.233$ & $0.1+0.16$ & $0.2+0.2$ & $2.0+3.0$ & $4+1$ & $3.2+1.8$ & $0.5+0.3$ \\
$10$ & $0.337$ & $1.258$ & $0.168$ & $0.203$ & $0.1+0.16$ & $0.2+0.2$ & $2.4+3.6$ & $4+1$ & $3.2+1.8$ & $0.5+0.3$ \\
\hline
All & - & - & - & - & $1+1.6$ & $1.4+1.4$& $13.2+19.8$ & $40+10$ & $32+18$ & $5.0+3.0$\\
\hline\hline
\end{tabular}
\label{tab:time}
\end{table*}

Table\,\ref{tab:time} summarizes the estimation of the hours required to follow up the 10 best-localized events in a year of observation with HLI\# for each facility. The sky area for the 10 best-localized events for HLV+ is also presented for comparison.
\begin{figure}
    \centering
    \includegraphics[width=1.00\linewidth]{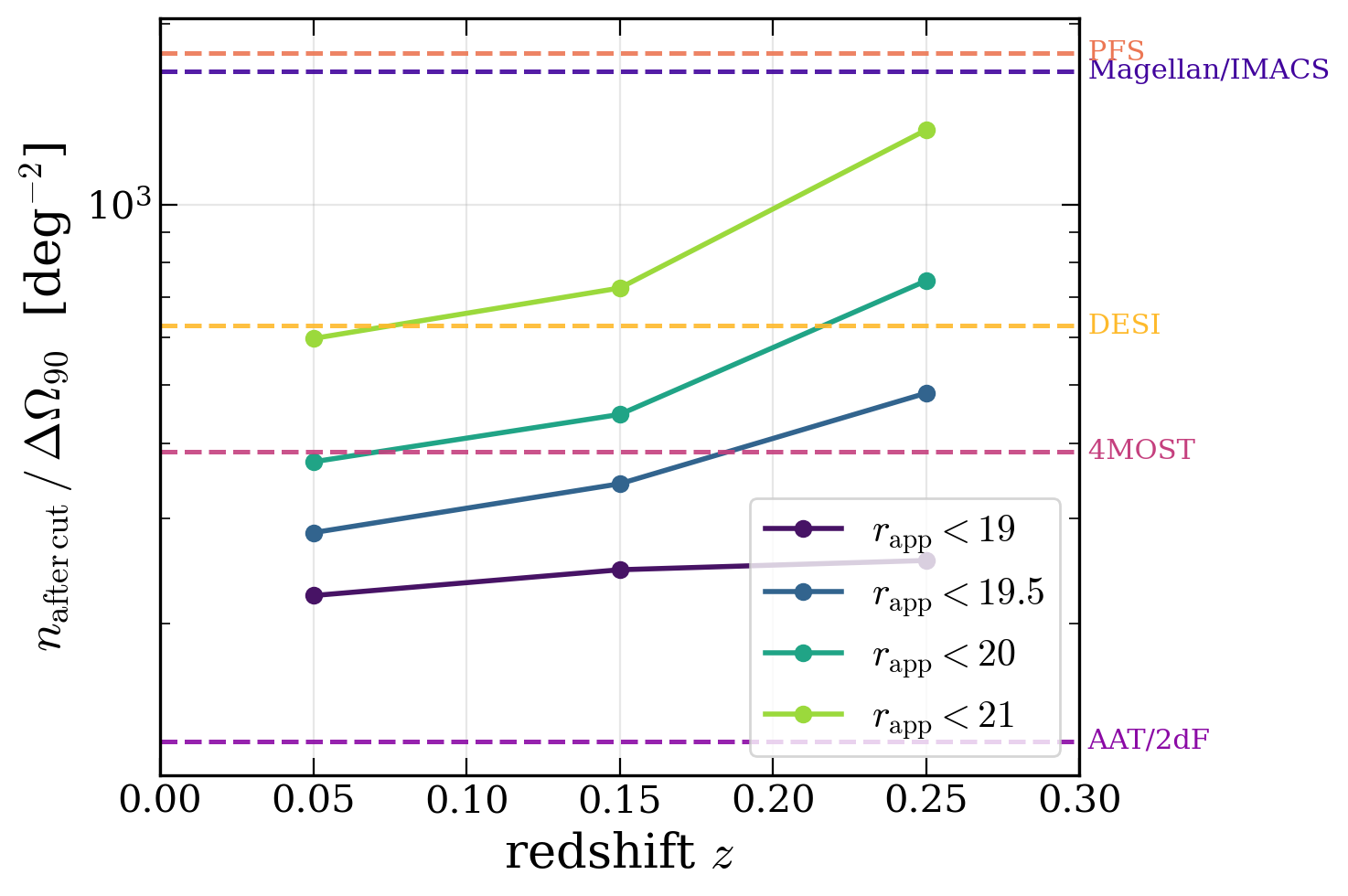}
    \caption{Mean target density, $n_\mathrm{after\,cut}/\Delta\Omega_{90}$, as a function of redshift for the $298$ best-localized events, where for each event the number of catalog galaxies surviving the apparent-magnitude cut within the $90\%$ credible region is divided by that region's sky area $\Delta\Omega_{90}$ and averaged in redshift bins of width $0.1$. Curves correspond to apparent-magnitude limits $r_\mathrm{app}<19,\,19.5,\,20,\,21$ (brighter cuts retain fewer galaxies and yield lower densities). Dashed horizontal lines mark the fiber/slit density of existing spectroscopic instruments: Magellan/IMACS ($1667\ \mathrm{deg}^{-2}$), AAT/2dF ($127\ \mathrm{deg}^{-2}$), PFS ($1794\ \mathrm{deg}^{-2}$), HET+VIRUS ($7.7\times10^{5}\ \mathrm{deg}^{-2}$), 4MOST ($387\ \mathrm{deg}^{-2}$) and DESI ($628\ \mathrm{deg}^{-2}$); a galaxy density below an instrument's line indicates its multiplexing capacity is sufficient to target all candidate hosts in a single pointing. HET+VIRUS is not shown in the plot as it far exceeds other data.}

    \label{fig:fiber density}
\end{figure}

\emph{Magellan/IMACS}---We adopt IMACS in f/2 multi-object spectroscopy mode on the 6.5\,m Magellan Baade telescope at Las Campanas Observatory. The f/2 camera delivers a circular field of view of 27.4\,arcmin diameter, corresponding to an effective sky coverage of $\Omega_\mathrm{FOV} \approx 0.18\,\mathrm{deg}^2$ per pointing with 5\% overlap between adjacent tiles, and accommodates $\sim$300 simultaneous slit spectra per mask with $R \sim 700$--$900$ over $\lambda \approx 400$--$900$\,nm. 

Based on the Las Campanas Observatory Exposure Time Calculator (LCO ETC) \citep{Blanc2016lcoetc} and assuming airmass 1.5, seeing 0.65\,arcsec, and 7\,days from new moon, a single 120\,s exposure achieves $S/N \approx 8$--$10$ for a source at $r = 20\,\mathrm{AB}$. We adopt $3\times120\,\mathrm{s}$ (6\,min on-sky) per mask to enable cosmic-ray rejection via spatial dithers. Following the IMACS User Manual \citep{Dressler2016imacs}, a field position accurate to a few arcseconds reduces the per-mask alignment time to 5--10\,min; we adopt 7\,min per mask as a representative value. For programs requiring more than 6 masks, replacing the full complement incurs an additional $\sim$30\,min overhead but this does not apply to cases considered here. The number of pointings required to tile a localization region of area $\Delta\Omega_{90}$ is
$N = \left\lceil \frac{\Delta\Omega_{90}}{\Omega_\mathrm{FOV}} \right\rceil= \left\lceil \frac{\Delta\Omega_{90}}{0.18\,\mathrm{deg}^2} \right\rceil$,
giving total on-sky time $T_\mathrm{exp} = N \times 6\,\mathrm{min}$ and total overhead $T_\mathrm{OH} = N \times 7\,\mathrm{min}$. Additional pointings might be needed depending on the shape of the estimated sky area and fiber collision due to the local overdensity. We assume no extra pointings are required as the slit density exceeds target by far as shown in Figure\,\ref{fig:fiber density}.

% For a golden dark siren with $\Delta\Omega_{90} \sim 0.1\,\mathrm{deg}^2$, a single IMACS pointing suffices ($N=1$), requiring 6\,min on-sky and 30\,min of mask overhead, for a total of $\sim$0.1\,hr\,+\,0.5\,hr. For a silver dark siren with $\Delta\Omega_{90} \sim 1\,\mathrm{deg}^2$, six pointings are required ($N=6$), giving 36\,min ($\sim$0.6\,hr) on-sky and 0.5\,hr of mask overhead, for a total of $\sim$0.6\,hr\,+\,0.5\,hr.

\emph{Anglo-Australian Telescope (AAT/2dF+AAOmega)}---The Anglo-Australian Telescope (AAT) is a 3.9\,m telescope at Siding Spring Observatory, Australia. For wide-field multi-object spectroscopy it is equipped with the 2dF fiber positioner \citep{AAT2df} feeding the AAOmega spectrograph \citep{Sharp2006}. The 2dF system deploys up to 400 science fibers over a 2\,degree diameter circular field of view % [Lewis+2002, Table 1, p.10]
corresponding to $\Omega_\mathrm{FOV} \approx 3\,\mathrm{deg}^2$ per pointing. AAOmega provides wavelength coverage 370--950\,nm at spectral resolution $R = 1000$--$8000$ \citep{Sharp2006}.

\citet{AAT2df} demonstrated that galaxies with $B \lesssim 19.5$ can be observed to $S/N \gtrsim 10$ per pixel in sets of $3 \times 1100$\,s ($\approx 55$\,min) integrations, % [Lewis+2002, Section 5.8, p.17: "3x1100-s integrations on sets
% of galaxies to B<19.5"]
with a redshift success rate of $\sim$95\%. % [Lewis+2002, abstract, p.1]
We adopt $T_\mathrm{exp} = 1$\,hr per pointing, appropriate for our $r = 19.5$ magnitude limit which is comparable to the 2dFGRS depth.

As shown in Figure\,\ref{fig:fiber density}, at $r=19.5$, the expected $\sim400~\mathrm{deg}^{-2}$ target density exceeds the fiber density, requiring conservatively 4 configurations of the same field; since reconfiguration runs in parallel with the preceding exposure, the total on-sky time scales accordingly to $\sim$4\,hr, with overhead remaining $\sim$1\,hr \citep{AAT2df}. 

\emph{4MOST}---The 4-metre Multi-Object Spectroscopic Telescope \citep[4MOST;][]{deJong2019} is a wide-field, high-multiplex fiber-fed spectroscopic survey facility mounted on the 4.1\,m VISTA telescope at ESO Paranal Observatory. Its focal plane has a hexagonal field of view of 4.2\,deg$^2$ and 2436 fibers with an on-sky aperture of 1.45\,arcsec diameter, with a minimum fiber separation of 15\,arcsec \citep{4MOSTCapabilities}. Of the 2436 fibers, 1624 feed two low-resolution spectrographs (LRS; $R = 4000$--$7700$, 370--950\,nm) and 812 feed a high-resolution spectrograph ($R = 18\,000$--$21\,000$). The LRS radial velocity accuracy is $<1$\,km\,s$^{-1}$, more than sufficient for dark siren host redshifts. Overheads are 3.5\,min for field acquisition and 4.4\,min per science exposure including attached calibrations
\citep{4MOSTCapabilities}.

The 4MOST LRS sensitivity curve \citep{4MOSTCapabilities} indicates that 90\% galaxy redshift completeness is achieved at $ \sim 20$\,AB in $6\times20$\,min ($= 2$\,hr) under new moon, mean seeing conditions. Applying a simple scaling for our target magnitude at $r=19.5$, the exposure time is estimated as 0.80\,hr. For a single configuration of $N_\mathrm{exp}$ exposures, the total overhead is $3.5 + N_\mathrm{exp} \times 4.4$\,min. Four separate exposures are conservatively considered as the the dark siren target density slightly exceeds the fiber density in the 0.2 to 0.3 redshift bin. Hence the time estimate is $\sim3.2$\,+\,1.8\,hr per event.

\emph{DESI}---The Dark Energy Spectroscopic Instrument (DESI) is a robotic, fiber-fed multi-object spectrograph on the 4\,m Mayall Telescope at Kitt Peak National Observatory \citep{DESI:2022xcl}. Its key specifications are a 3.2\,degree diameter field of view ($\Omega_\mathrm{FOV} \approx 8.0$\,deg$^2$)
% [Silber+2022: "3 degree diameter field of view"]
and 5000 robotically positioned science fibers feeding ten three-arm spectrographs, covering 360--980\,nm at resolution $R = 2000$--$5500$. The fiber positioners reconfigure sequentially: all 5020 fibers are re-targeted between exposures with a reconfiguration time of under 2\,min,
% [Silber+2022: "reconfiguration time less than 2 minutes"]
after which the next exposure begins. 

The DESI Bright Galaxy Survey (BGS) defines a nominal exposure time of $t_\mathrm{nom} = 180$\,s, calibrated to achieve $>95\%$ redshift success
% [Hahn+2023, Figure 5: "overall redshift success rate of ~95% for r<19.5
% with tnom=180s"]
for an $r < 19.5$\,AB magnitude-limited sample under nominal dark conditions (zenith, seeing $1.1''$, no extinction). Exposure times are dynamically scaled by the DESI ETC to account for varying observing conditions \citep{Hahn2023}. We adopt $T_\mathrm{exp} = 180$\,s ($= 0.05$\,hr) per pointing. The dominant overhead per pointing is the fiber reconfiguration plus readout, which is $\lesssim 2$\,min, plus a fixed cost of $\sim 3$\,min for telescope slewing and acquisition. We adopt a total overhead of $T_\mathrm{OH} \approx 5$\,min ($\approx 0.08$\,hr) per pointing. In Figure\,\ref{fig:fiber density}, we observe that the target density at $r\leq20$ can be close to the fiber density of DESI. After running fiber assignment, we conclude 2 passes per event on average for a conservative estimate.

% Both a golden dark siren ($\Delta\Omega_{90} \sim 0.1$\,deg$^2$) and a silver dark siren ($\Delta\Omega_{90} \sim 1$\,deg$^2$) fit comfortably within a single DESI pointing. For the GDS, the $\sim$80 galaxy candidates are far within the 5020-fiber capacity: a single 180\,s configuration suffices, giving $T_\mathrm{exp} = 0.05$\,hr and $T_\mathrm{OH} = 0.08$\,hr. For the SDS, the $\sim$800 candidates likewise fit within one configuration ($800 \ll 5020$), giving identical estimates: $T_\mathrm{exp} = 0.05$\,hr and $T_\mathrm{OH} = 0.08$\,hr. In both cases the overhead exceeds the exposure time, reflecting the exceptionally shallow depth required at $r = 19.5$.

\emph{Subaru/PFS}---The Prime Focus Spectrograph \citep[PFS;][]{Takada2014} is a massively multiplexed, wide-field optical and near-infrared multi-object spectrograph on the 8.2\,m Subaru Telescope at Maunakea, Hawaii, which began open-use  science operations in February 2025. PFS deploys 2386 science fibers over a 1.25\,degree diameter field of view \citep{SubaruPFSInstrument}.

Queue-mode observations with PFS are constrained to a minimum single-frame exposure time of 7.5\,min, and we therefore adopt $T_\mathrm{exp} =4\times 7.5$\,min ($= 0.5$\,hr) per configuration as the practical minimum, which is sufficient for galaxies at low redshift at our target apparent magnitude.
% [PFS Observations page: "The exposure time for a single frame is fixed
% to 15 min."]
We adopt the overhead time of $\approx 0.15$\,hr per pointing. With 2386 fibers over 1.33\,deg$^2$, the effective fiber density is $\sim$1794\,fibers\,deg$^{-2}$, far exceeding the dark siren target density as shown in Figure\,\ref{fig:fiber density}. We stress that the pointing simulation constrains only fiber allocation and observing time, not the redshift success rate and additional exposures may therefore be needed.

% A single configuration suffices for all best 10 localized events without any need for multiple passes. The time estimate per event is therefore a total of $\sim$0.5\,+\,1.28\,hr per event.

\emph{HET/VIRUS}---The Hobby--Eberly Telescope \citep[HET;][]{Ramsey1998} is a 10\,m fixed-altitude telescope at McDonald Observatory, Texas, equipped with the Visible Integral-field Replicable Unit Spectrograph \citep[VIRUS;][]{het_descrip}. VIRUS consists of 78 identical IFUs, each covering a 51$\times$51\,arcsec field with 448 fibers of 1.5\,arcsec diameter, feeding 156 spectrograph channels covering 350--500\,nm\ at $R \sim 750$--$950$ \citep{het_descrip,Hill2021}. The IFUs are arrayed on a 100\,arcsec center-to-center grid, leaving 49\,arcsec gaps, so that a single dithered observation covers $\sim$1/4.5 of the sky within the instrument footprint \citep{Hill2021}. % a single exposure covers only ~1/3 of this before the dither completes
Complete contiguous sky coverage requires two levels of multiplexing. First, a \textbf{3-point dither} shifts the telescope by $\sim$1.5\,arcsec between exposures to fill the intra-IFU fiber gaps, achieving complete coverage within each 51$\times$51\,arcsec IFU. Second, a \textbf{4-point tiling} shifts the entire instrument by $\sim$50\,arcsec (half the IFU pitch) between dithered observations to fill the inter-IFU gaps.

\citet{Hill2021} report S/N $=10$ at $g=19.5$ in a median 1800\,s exposure. We adopt $3\times2$\,min $= 6$\,min per tiling position. The total on-sky time per tiled set is therefore $4\times6$\,min $= 24$\,min.

We adopt the tiling overhead model from the HET observing documentation \cite{HET_VIRUS}, which estimates $\sim$35\,min of total overhead for a complete four-position VIRUS map (three dithers per position), comprising target setup at each tiling position and CCD readout after each dither. This estimate is conservative and consistent with the per-step values reported for the operational survey \citep{Hill2021, het_descrip}.

%% ── Bibliography ──────────────────────────────────────────────────────────────
\bibliography{references}
\bibliographystyle{unsrtnat} 

\end{document}